\documentclass[aps,prl,twocolumn,floatfix,superscriptaddress,showpacs,amsfonts,amssymb,amsmath,preprintnumbers]{revtex4}
\usepackage{bm}
\usepackage{psfig}
\usepackage{epsfig}
\usepackage{color}

\newcommand{\eq}[1]{Eq.~(\ref{#1})}

\newcommand{\fig}[1]{Fig.~\ref{#1}}
\newcommand{\Fig}[1]{Fig.~\ref{#1}}

\newcommand{\etal}{\textit{et al.}}
\newcommand{\const}{{\rm const}}

\newcommand{\bea}{\begin{eqnarray}}
\newcommand{\eea}{\end{eqnarray}}
\newcommand\be{\begin{equation}}
\newcommand\ee{\end{equation}}

\newcommand{\jpar}{j_\parallel}
 
\newcommand{\Bperp}{\bm{B}_\perp}

\newcommand{\Vperp}{\bm{v}_\perp}

\newcommand{\geff}{\gamma_\text{eff}}
\newcommand{\gcoll}{\gamma_\text{SP}}

\newcommand{\Wc}{W_\text{c}}

\newcommand{\Wsat}{W_\text{sat}}
\newcommand{\psix}{\Psi}
\newcommand{\psic}{\Psi_\text{c}}
\newcommand{\psisat}{\Psi_\text{sat}}

\newcommand{\Bin}{B_\text{in}}
\newcommand{\vin}{v_\text{in}}
\newcommand{\vout}{v_\text{out}}
\newcommand{\Bo}{B_0}
\newcommand{\tc}{t_\text{c}}
\newcommand{\Lsheet}{L_\text{CS}}
\newcommand{\deltacs}{\delta_\text{CS}}

\newcommand{\ez}{\bm{e}_z}
\newcommand{\bnab}{\bm{\nabla}}
\newcommand{\nabperp}{\bnab}

\newcommand{\DD}{\Delta'}

\renewcommand{\d}{\partial}

\newcommand{\REM}[1]{\textcolor{red}{\bf #1}}

\begin{document}


\title{$X$-point collapse and saturation in the nonlinear tearing mode reconnection}
\author{N.\ F.\ Loureiro}
\email{n.loureiro@imperial.ac.uk}
\affiliation{Plasma Physics Group, Imperial College, 
Blackett Laboratory, Prince Consort Road, London~SW7~2BW, UK}
\author{S.\ C.\ Cowley}
\affiliation{Plasma Physics Group, Imperial College, 
Blackett Laboratory, Prince Consort Road, London~SW7~2BW, UK}
\affiliation{Department of Physics and Astronomy, 
UCLA, Los Angeles, California 90095-1547}
\author{W.\ D.\ Dorland}
\affiliation{Department of Physics, University of Maryland, College Park, Maryland 20742-3511} 
\author{M.\ G.\ Haines}
\affiliation{Plasma Physics Group, Imperial College, 
Blackett Laboratory, Prince Consort Road, London~SW7~2BW, UK}
\author{A.\ A.\ Schekochihin}
\affiliation{DAMTP, University of Cambridge, Cambridge CB3 0WA, UK}
\date{\today}

\begin{abstract}
We study the nonlinear evolution of the resistive tearing mode in slab geometry in two dimensions. 
We show that, in the strongly driven regime (large $\DD$), a collapse of the $X$-point occurs 
once the island width exceeds a certain critical value $\sim 1/\DD$. A current sheet is formed  
and the reconnection is exponential in time with a 
growth rate $\propto\eta^{1/2}$, where $\eta$ is the resistivity. 
If the aspect ratio of the current sheet is sufficiently large, 
the sheet can itself become tearing-mode unstable, giving rise to
secondary islands, which then coalesce with the original island. 
The saturated state depends on the value of $\DD$. 
For small $\DD$, the saturation amplitude is $\propto\DD$ and 
quantitatively agrees with the theoretical prediction. 
If $\DD$ is large enough for the $X$-point collapse to have 
occured, the saturation amplitude 
increases noticeably and becomes independent of $\DD$.
\end{abstract}

\pacs{52.35.Vd, 52.65.Kj, 52.35.Py}

\maketitle

\begin{figure}[b!]
\unitlength1cm
\psfig{file=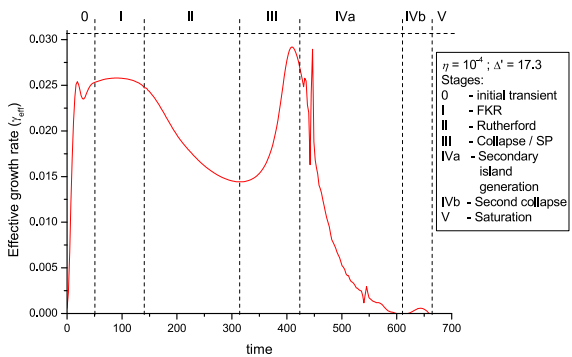,width=7.5cm}
\caption{\label{intro_figs} 
Effective growth rate at the $X$-point $\geff=d\ln\psix/dt$ 
vs.~time for a strongly driven (large $\DD$) tearing mode.}
\end{figure}

Magnetic reconnection is the breaking and rejoining of magnetic field lines 
in a plasma. 
Solar flares are believed to be a manifestation of this phenomenon 
\cite{sweet_review}. 
Other classical examples are reconnection between the solar and the Earth's magnetic field 
in the magnetopause and the magnetotail \cite{dungey_61} 
and the sawtooth instability in tokamaks \cite{kadomtsev}. 
In some cases, most notably the sawtooth, 
reconnection takes place in a plane perpendicular to a strong magnetic 
field, in which case it occurs via the tearing-mode instability. 
Linear theory \cite{FKR} shows that an MHD equilibrium is tearing-mode unstable if the 
instability parameter $\DD>0$. Analytical and numerical studies of the tearing mode have been 
mostly restricted to low values of $\DD$. However, it has been shown that kinetic effects can 
change the instability threshold to $\DD>\DD_\text{crit}\gg1$ \cite{cowley_86,migliuolo_91} 
and there is, 
indeed, experimental evidence for $\DD\gg1$ in the sawtooth \cite{GP}. 
The evolution of large-$\DD$ (i.e., strongly driven) tearing modes, even in the simplest 
physical models, remains poorly understood.
To address this problem, we investigate the evolution of the tearing mode in 
the broadest ranges of $\DD$ and the resistivity $\eta$ achieved to date. 
We find that, for sufficiently large $\DD$ and sufficiently small $\eta$, 
the tearing mode goes through five stages (\fig{intro_figs}): 
(I) linear instability \cite{FKR}, 
(II) algebraic growth (Rutherford \cite{ruth_73} stage), 
(III) $X$-point collapse followed by current-sheet reconnection 
(Sweet-Parker \cite{sweet_58a,parker} stage), 
(IV) tearing instability of the current sheet resulting in 
generation of secondary islands, 
and (V) saturation. 
The traditional theory of the tearing mode, valid for small $\DD$, 
comprises just Stages I, II and V. The fact that, at large $\DD$, 
Stage II is succeeded by Stages III, IV, and a modified Stage V 
is new in the tearing-mode context. 

We solve the Reduced MHD equations \cite{strauss_76} 
\bea
\label{RMHD_vort}
\d_t\omega + \Vperp \cdot \nabperp \omega &=&
\Bperp \cdot \nabperp \jpar,\\
\label{RMHD_psi}
\d_t\psi + \Vperp \cdot \nabperp \psi &=&
\eta \nabla^2 \psi
\eea
in a two-dimensional periodic box $L_x\times L_y$ 
using a pseudo-spectral code at resolutions 
up to $3072 \times 4096$. 
The total magnetic field is $\bm B = B_z\ez+\Bperp$, 
the in-plane magnetic field is $\Bperp = \ez\times\bnab\psi$, 
the in-plane velocity is $\Vperp = \ez\times\bnab\phi$, 
and $\omega = \ez\cdot(\bnab\times \Vperp) = \nabla^2\phi$,
$\jpar= \ez\cdot(\bnab\times\bm B) = \nabla^2\psi$.
We impose the equilibrium configuration 
$\psi^{(0)}=\psi_0/\cosh^2(x)$ and $\phi^{(0)}=0$. 
By setting $\psi_0=3\sqrt{3}/4$, we scale 
the units of field strength in such a way that the maximum value of 
$B_y^{(0)}=d\psi^{(0)}/dx$ is $B_{y,{\rm max}}^{(0)}=1$. 
All lengths are scaled so that $L_x=2\pi$. 
Time is, therefore, scaled by the in-plane Alfv\'en time 
$L_x/2\pi B_{y,{\rm max}}^{(0)}$.
To the equilibrium, we add an initial perturbation 
$\psi^{(1)}=\psi_{1}(x)\cos(ky)$, where $k=L_x/L_y$. 
Given a perturbation in this form, the island width 
$W$ and the reconnected flux $\psix(t)=-\psi(t,0,0)+\psi_0$ 
are related by 
\bea
\label{W_vs_psi}
W=4\sqrt{\psix(t)/\psi_0''(0)}. 
\eea

For our equilibrium, the instability parameter is \cite{porc_rec}
\bea
\label{Dprime_def}
\Delta' = \frac{\psi'_1(+0)-\psi'_1(-0)}{\psi_1(0)} 
= \frac{2(5-k^2)(3+k^2)}{k^2\sqrt{4+k^2}}.
\eea
The equilibrium is tearing-unstable if $\DD>0\Leftrightarrow k<\sqrt 5$. 
$\DD$ is varied by changing $k$, i.e., $L_y$.

We now describe the evolution of the tearing mode stage by stage. 
During Stages I--II, reconnection occurs via an $X$-point 
configuration. In Stage I, it is a linear instability 
with the island width $W$ growing exponentially in time \cite{FKR}. 
Once $W$ exceeds the resistive scale, 
this stage gives way to the Rutherford \cite{ruth_73} stage 
(Stage II), during which the growth is algebraic in time: $dW/dt\sim\eta\DD$. 
Omitting further discussion of these stages, which 
have been studied before \cite{biskamp_NMHD}, 
we proceed~to 

\begin{figure}[t]
\begin{tabular}{ccc}
\psfig{file=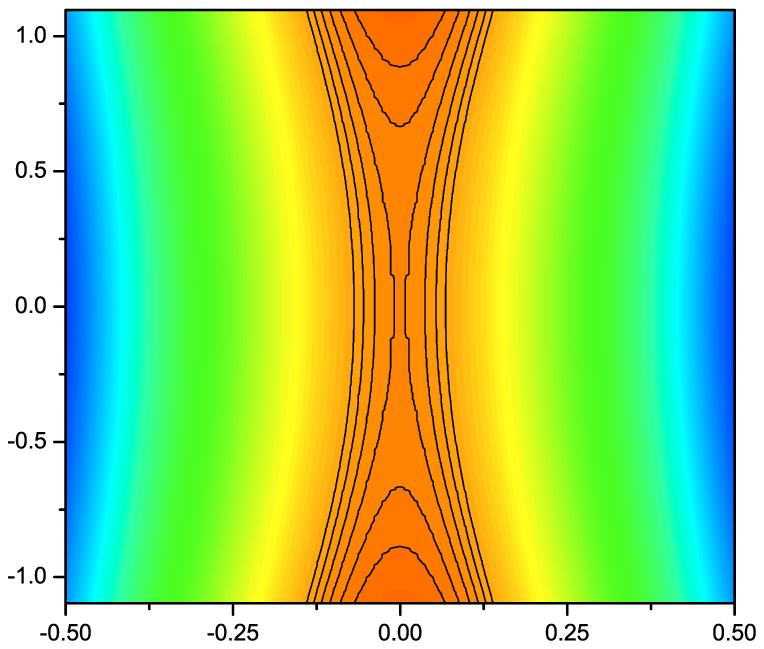,width=3.5cm} 
& \qquad\qquad &
\psfig{file=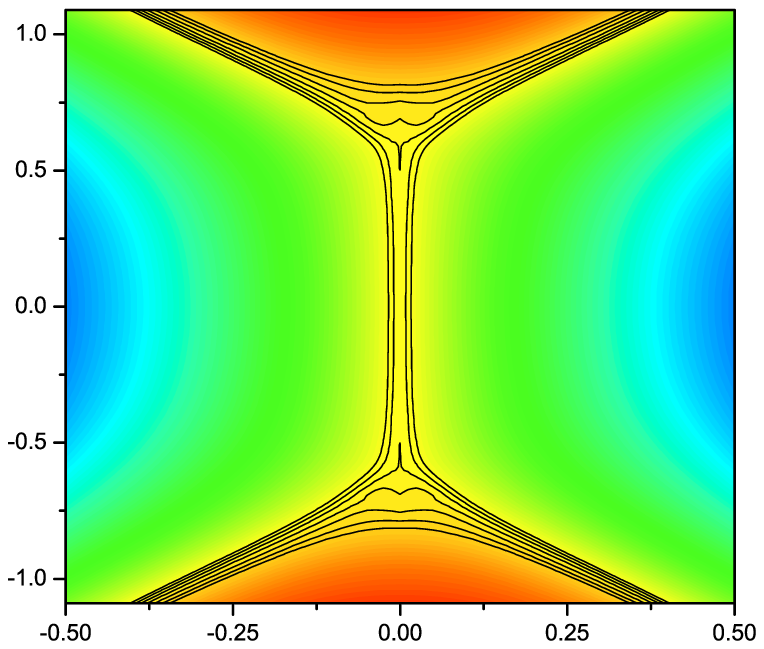,width=3.5cm}\\
(a) $t=314$ & & (b) $t=440$
\end{tabular}
\caption{\label{coll_contours} 
Contours of $\psi$ at the beginning and end of Stage III in \fig{intro_figs}. 
The boundaries of these plots are 
{\it not} the boundaries of the computational box.}
\end{figure}

\begin{figure}[t]
\psfig{file=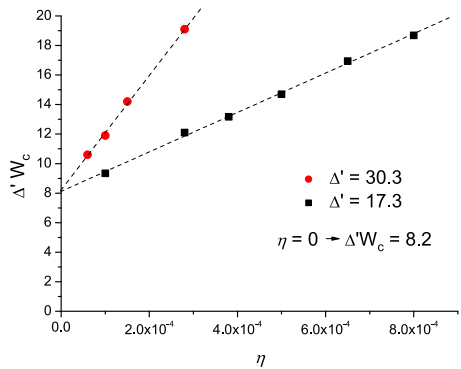,width=6cm}
\caption{\label{waelb_scaling} 
The critical island width for collapse vs.~$\eta$ 
at fixed $\DD=17.3,~30.1$. 
Dashed lines are linear fits.}
\end{figure}

\paragraph{Stage III: $X$-Point Collapse and Sweet--Parker Reconnection.}
In simulations with large $\DD$, the $X$-point 
configuration eventually collapses and a current sheet is formed 
(\fig{coll_contours}) 
accompanied by a dramatic speed up of the island growth (\fig{intro_figs}). 
Several previous numerical studies in various reconnection contexts 
have also reported a nonlinear speed-up \cite{aydemir_97} 
and a tendency for the current-sheet formation \cite{jemella_03}. 
Waelbroeck \cite{waelb_93} predicted that 
when the island width $W>\Wc\sim 1/\DD$, no equilibrium 
$X$-point configuration exists and a current sheet must form. 
By varying $\eta$ and $\DD$ in our simulations, we have tried to verify 
this prediction. We define $\Wc$ as the island width at which 
$d\geff/dt=0$ after the Rutherford-like algebraic stage (e.g., 
at $t\approx315$ in \fig{intro_figs}). 
In \fig{waelb_scaling}, we plot $\DD \Wc$ 
vs.~$\eta$ for two different values of $\DD$. 
The dependence of $\DD\Wc$ on $\eta$ appears to be linear 
and extrapolates in the limit of $\eta\to0$ to $\DD \Wc\simeq8.2$ 
for both values of $\DD$ used. Thus, the transition criterion is 
\bea
\label{Waelbroek}
\DD \Wc\simeq8.2+f(\DD)\eta,
\eea 
where the slope function $f(\DD)$ remains undetermined but is seen 
in \fig{waelb_scaling} to increase with $\DD$.

\begin{figure}[b]
\begin{tabular}{cc}
\psfig{file=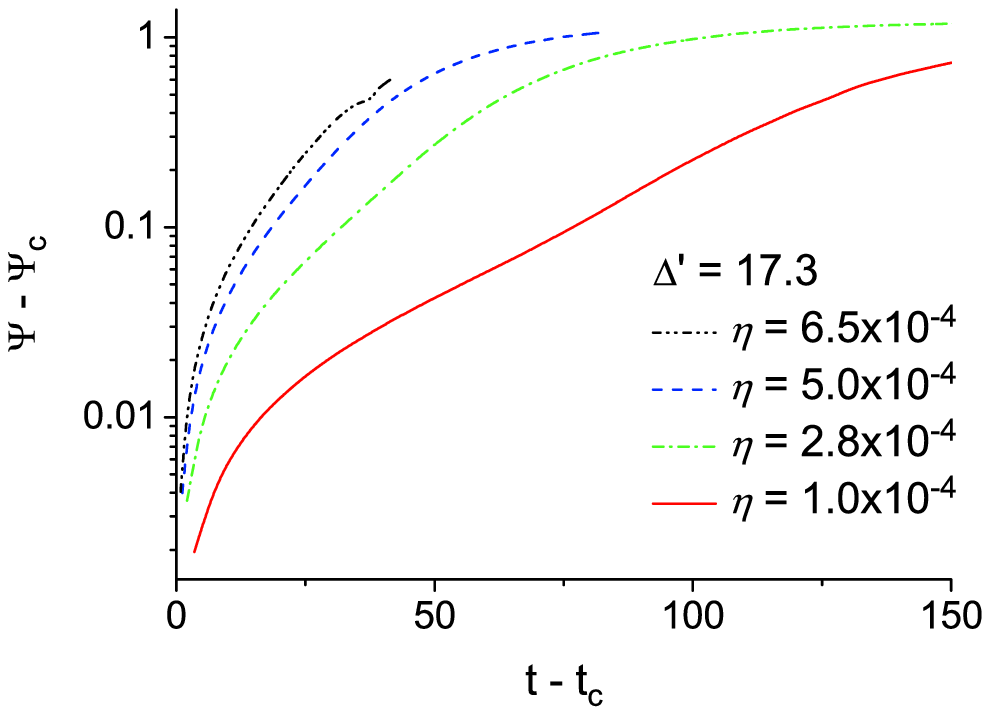,width=4.2cm} &
\psfig{file=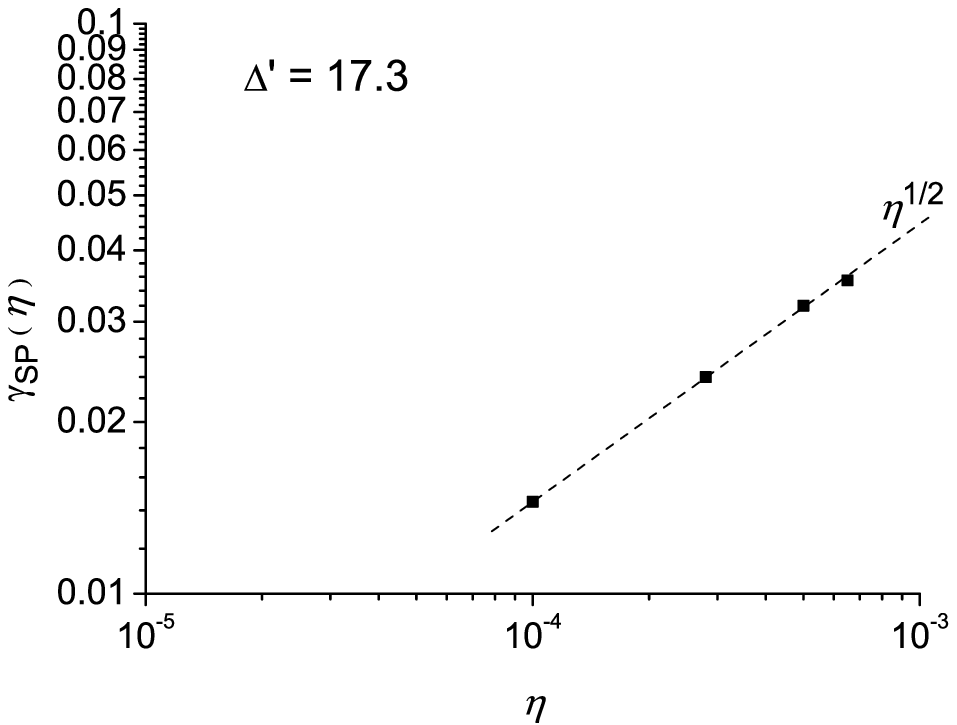,width=4.2cm}\\
(a) & (b)
\end{tabular}
\caption{\label{collapse_psi}
(a) Growth of the reconnected flux $\psix$ during the SP stage 
for fixed $\DD=17.3$ and various values of $\eta$.
(b) Slopes of these lines vs.~$\eta$ during the exponential growth.}
\end{figure}

\Fig{collapse_psi}(a) shows that, in this stage, the reconnected flux 
(measured at $x=y=0$) grows exponentially in time: $\ln(\psix-\psic)=\gcoll(t-\tc)$,
where $\tc$ is the time at which the collapse begins, $\psic=\psix(\tc)$, and 
$\gcoll$ is the growth rate  
\footnote{The exponential growth is not apparent in the $\geff$ diagnostic 
of \fig{intro_figs} because in our simulations, $\psix$ only grows by a factor of 
$\lesssim10$ during this stage.}.  
Varying $\DD$, we have ascertained that $\gcoll$ is independent of $\DD$. 
Its dependence on $\eta$ is plotted in \fig{collapse_psi}(b). 
The scaling $\gcoll\propto\eta^{1/2}$ appears to hold. 

\begin{figure}
\begin{tabular}{cc}
\psfig{file=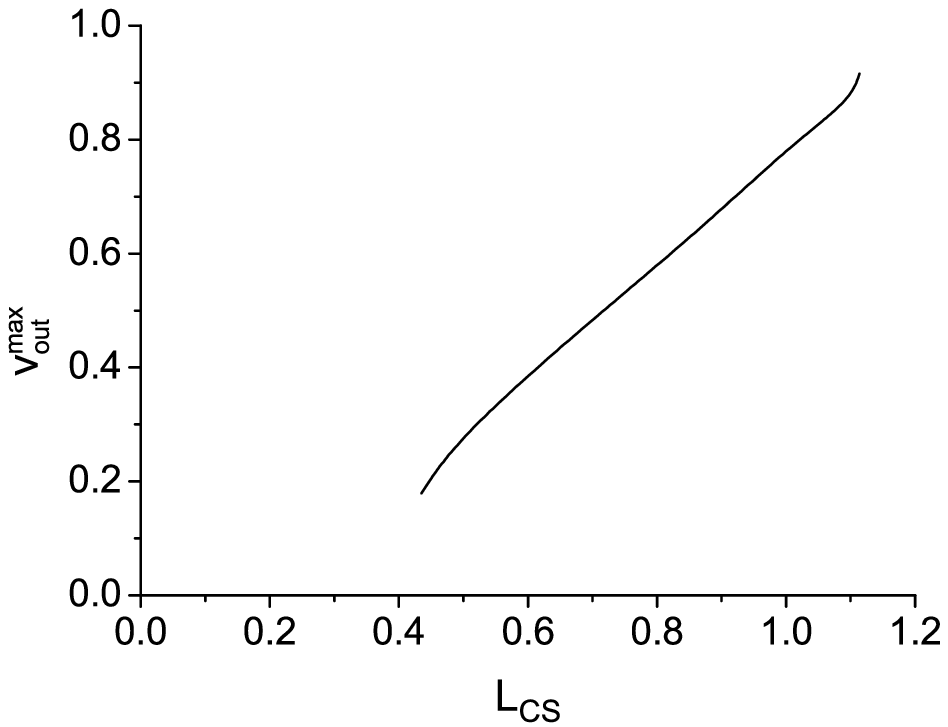,width=4.2cm} &
\psfig{file=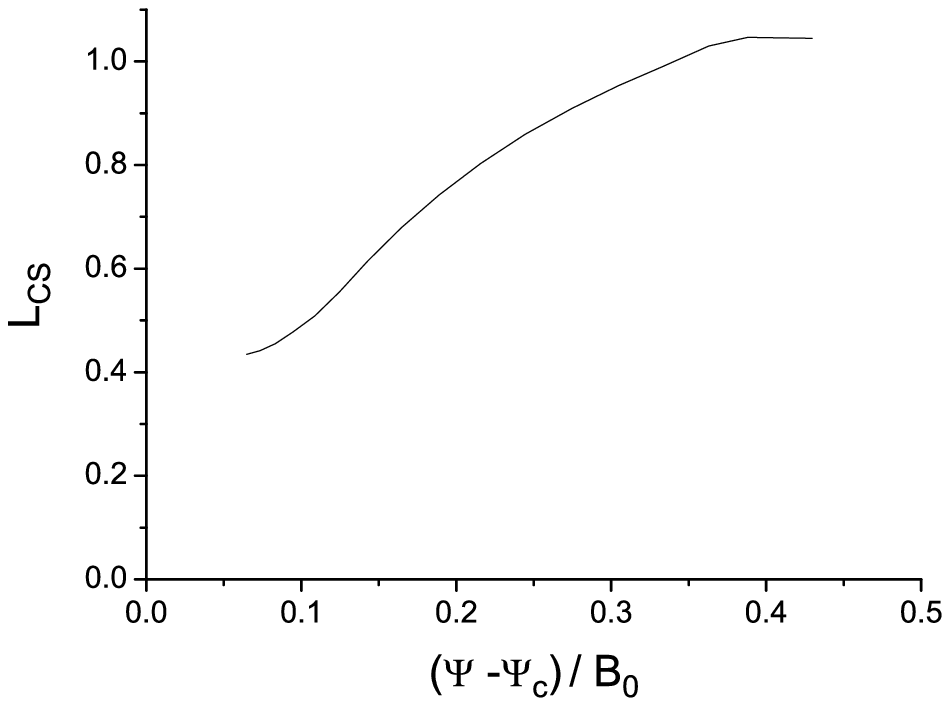,width=4.2cm}\\
(a) & (b) 
\end{tabular}
\caption{\label{out_vs_length} 
The exponential stage ($370<t<450$) of the run of \fig{intro_figs}: 
(a) maximum outflow velocity $\vout$ vs.~the current-sheet length 
$\Lsheet$; 
(b) $\Lsheet$ vs.\ $(\psix-\psic)/\Bo$, where 
$\Bo$ is defined as the maximum value of $B_y$ along the $x$ axis. 
These curves do not extrapolate to the origin because the full-width-half-maximum 
definition used for $\Lsheet$ correctly reflects the growth of the current-sheet
length but not its true length (thus, it formally gives $\Lsheet>0$ for 
the $X$-point reconnection).}
\end{figure}

We think that what we observe is an exponential-in-time Sweet-Parker (SP) reconnection 
that proceeds qualitatively in the way described in \cite{sweet_58a,parker} 
but with the outflow velocity $\vout$ and the current sheet length $\Lsheet$ growing with 
time. Since the reconnected flux $\psix$ changes at the SP rate $\propto\eta^{1/2}$, 
we can assume that the evolution is quasistatic, so that 
the system passes through a sequence of ideal equilibria, 
in each of which $\Lsheet$ and 
the configuration outside (but not inside) the current sheet 
are fully determined by the instantaneous value of $\psix$. 
Let us assume that in these equilibria, the vicinity 
of the current sheet is described by the 
Syrovatskii solution with a unidirectional current \cite{syro_71}. 
In this solution, the magnetic field immediately outside the sheet is 
$\Bin=\Bo(\psix)\Lsheet(\psix)/L_x$, where $\psix=\psix(t)$ is the 
reconnected flux and $\Bo$ is the field away from the sheet. 
Then the reconnected flux grows according to 
(cf.~\cite{waelb_89,jemella_03})
\bea
{d\psix\over dt}\sim\vin\Bin\sim\eta^{1/2}\left[{\Bo(\psix)\over L_x}\right]^{3/2} 
\Lsheet(\psix),
\label{psi_eq}
\eea
where we have used the SP expression for the inflow velocity, 
$\vin\sim(\eta\vout/\Lsheet)^{1/2}$, and  
taken the outflow velocity to be Alfv\'enic, $\vout\sim\Bin$. 
\eq{psi_eq} implies that the growth of $\psix$ must speed up compared 
to $\psix\propto(\eta\DD t)^2$ in the Rutherford \cite{ruth_73}
stage (Stage II). 
When $\psix$ is close to its value $\psic$ 
at the beginning of the collapse, we may approximate 
$\Bo(\psix)\sim\Bo(\psic)=\const$. This implies 
$\vout/\Lsheet\sim\Bin/\Lsheet\sim\Bo/L_x=\const$, 
a conclusion confirmed by \fig{out_vs_length}(a). 
Since $\Lsheet(\psic)=0$, $\Lsheet$ should be a growing function of 
$\psix-\psic$. Indeed, \eq{psi_eq} is consistent with the numerically 
observed exponential SP reconnection if $\Lsheet\sim (\psix-\psic)/\Bo$ 
[cf.\ \fig{out_vs_length}(b)]. 

\begin{figure}[b]
\begin{tabular}{cc}
\psfig{file=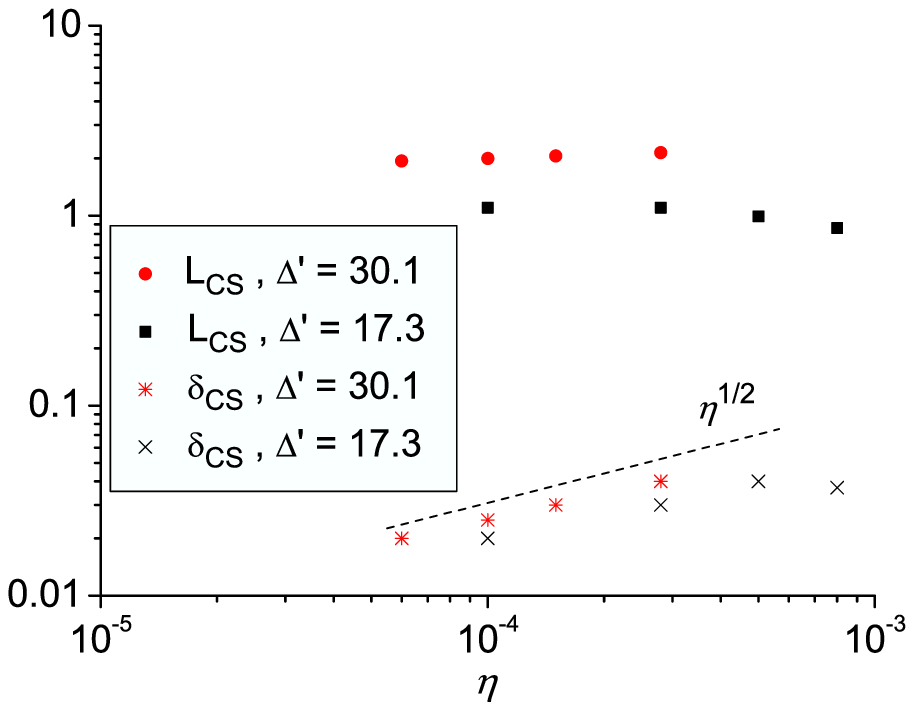,width=4.2cm} &
\psfig{file=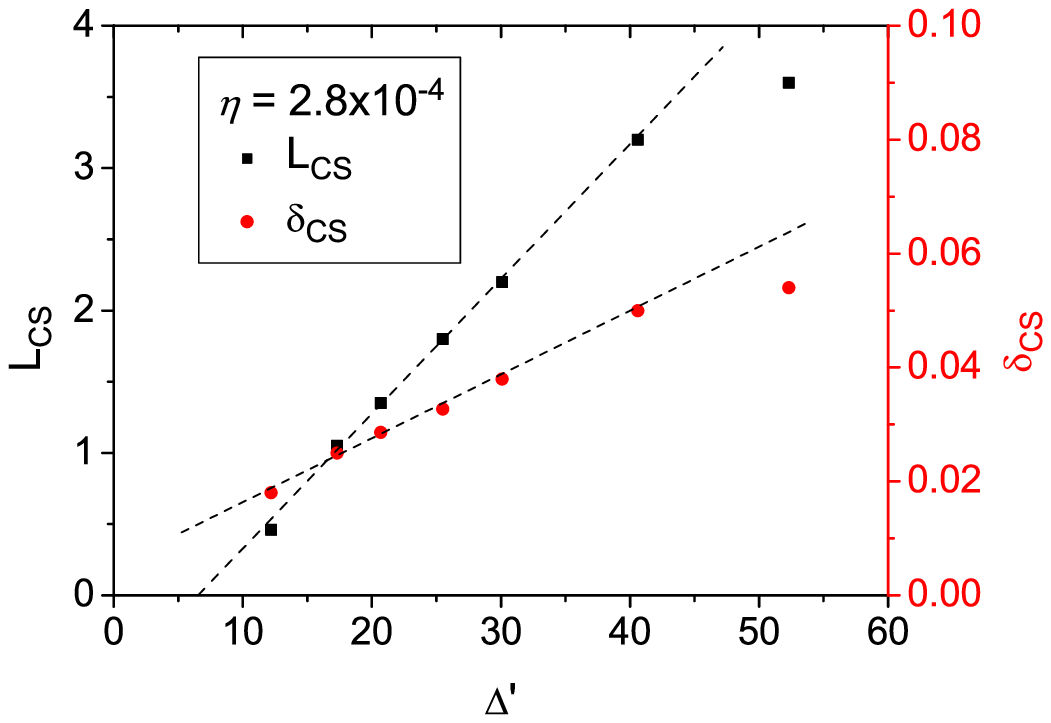,width=4.2cm}\\
(a) & (b)
\end{tabular}
\caption{\label{length_width}
The current sheet length $\Lsheet$ and width $\deltacs$ 
(a) vs.~$\eta$ and (b) vs.~$\DD$.}
\end{figure}

The elongation of the current sheet ceases when $\Lsheet$ reaches a 
significant fraction of the box size. Reconnection can still proceed 
in a SP fashion, but the growth of the reconnected flux slows down 
(see \fig{intro_figs}). Indeed, in the right-hand side of \eq{psi_eq}, 
$\Lsheet$ no longer increases with $\psix$ 
and $\Bo(\psix)$ starts to decrease as the initial 
reconnectable flux is used up. 
In \fig{length_width}, 
we show the current-sheet length $\Lsheet$ and width $\deltacs$ 
measured using a full-width-half-maximum estimate at the time 
when the maximum $\Lsheet$ is reached. 
We see that, for fixed $\DD$, $\Lsheet$ is roughly independent of $\eta$, 
while $\deltacs\sim\eta^{1/2}$, in agreement with the SP prediction.
On the other hand, for fixed $\eta$, both $\Lsheet$ and $\deltacs$ grow 
linearly with $\DD$ (cf.~\cite{jemella_04}), 
except for the largest data point, $\DD=52.7$ 
\footnote{The deviation from linearity is a finite-box-size effect. 
For $\DD\gg1$, we have $\DD\simeq15/k^2\propto L_y^2$ [see \eq{Dprime_def}]. 
Since $\Lsheet$ cannot exceed the box length $L_y$, 
it must, at large $\DD$, grow slower that $\sqrt{\DD}$.}.



\paragraph{Stage IV: Secondary Island Generation.}

When the aspect ratio of the current sheet 
$A=\Lsheet/\deltacs\gtrsim50$, the sheet itself becomes unstable to tearing 
modes and generates secondary islands. 
We expect that this critical value is independent of either $\DD$ 
or $\eta$, but due to resolution constraints, we do not yet 
have a numerical confirmation of this conjecture 
\footnote{In forced-reconnection simulations, a destabilization of the current 
sheet has also been seen at $A\sim10^2$, though opinions on whether 
this value depends on $\eta$ vary \cite{biskamp_NMHD,lee_fu}.}. 

\begin{figure}[t]
\begin{tabular}{ccc}
\psfig{file=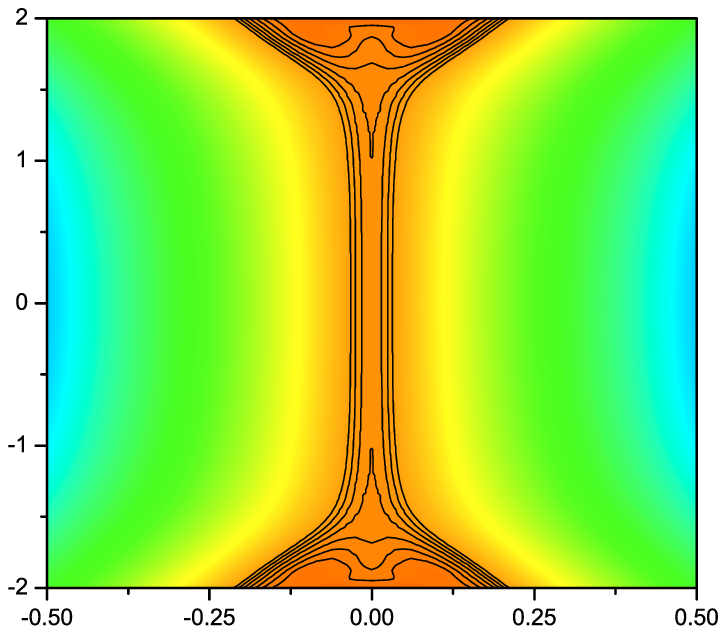,width=2.75cm} &
\psfig{file=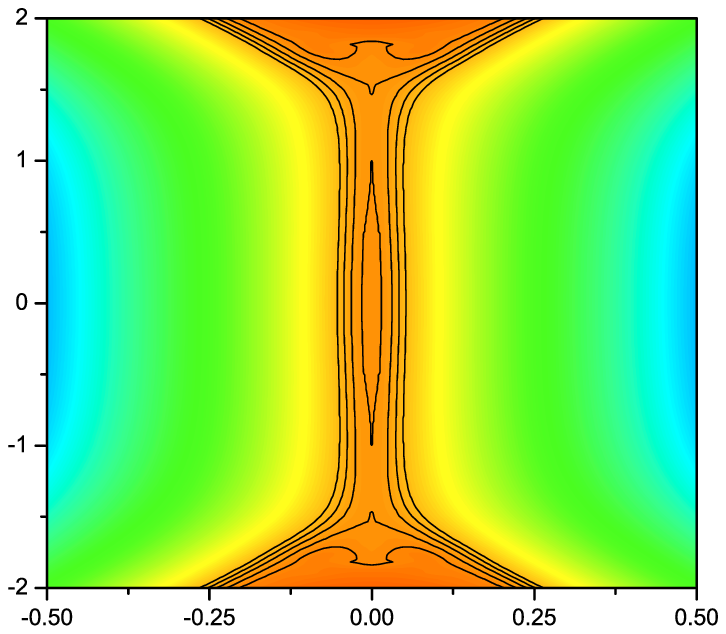,width=2.75cm} &
\psfig{file=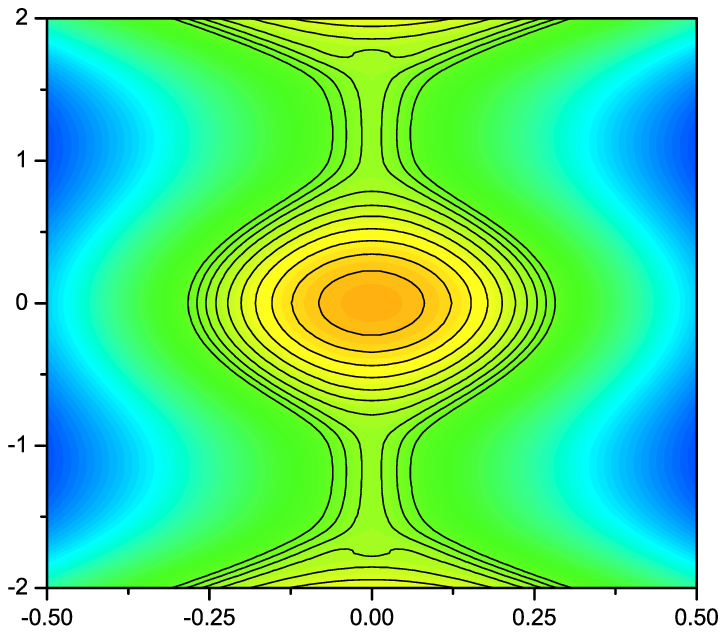,width=2.75cm} \\
(a) & (b) & (c) \\\\
\psfig{file=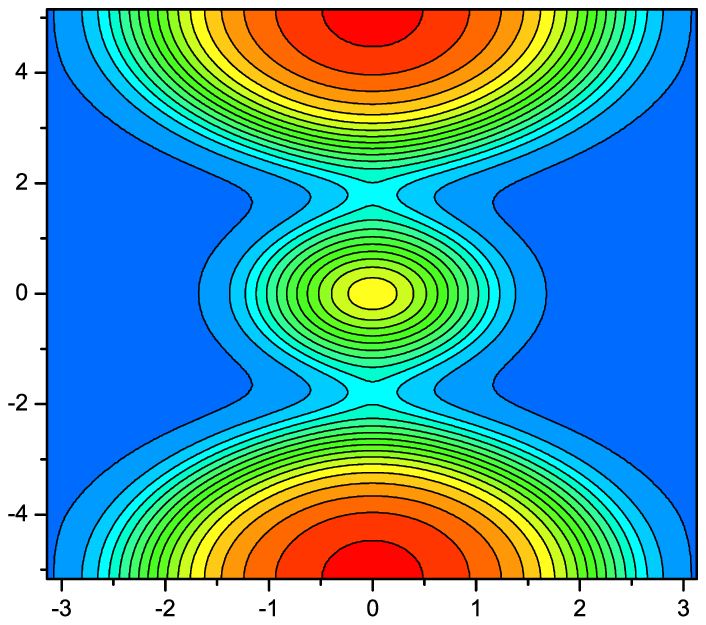,width=2.75cm} &
\psfig{file=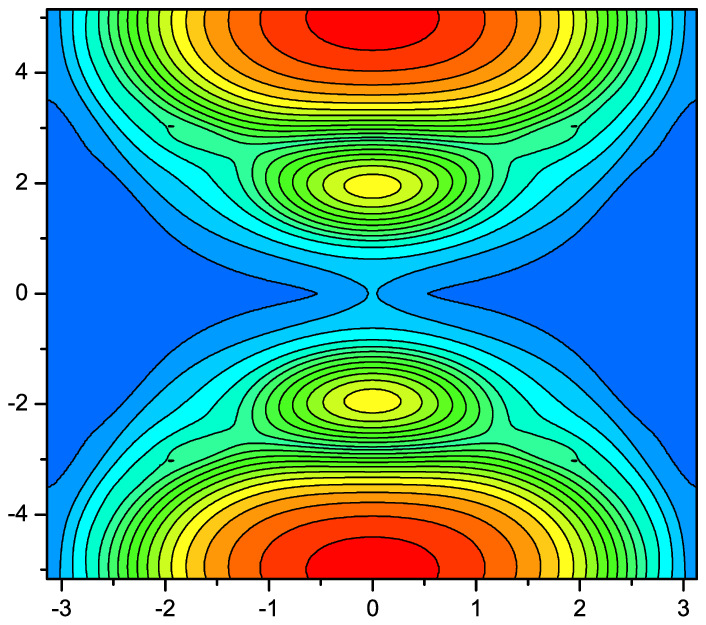,width=2.75cm} &
\psfig{file=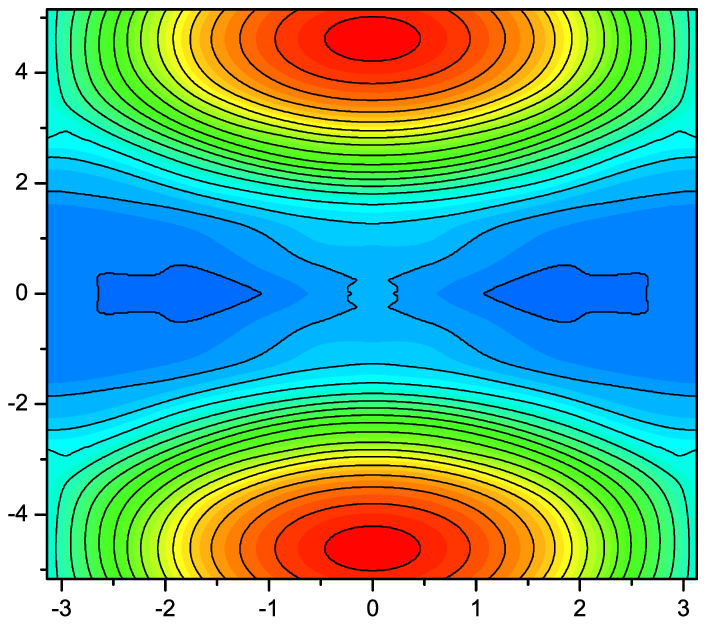,width=2.75cm}\\
(d) & (e) & (f)
\end{tabular}
\caption{\label{plasmoid} 
Contours of $\psi$ showing the current sheet instability 
(a-c) and the subsequent nonlinear evolution of the secondary island
(d-f) for a run with $\DD=40.6,~\eta=2.8\times10^{-4}$.} 
\end{figure}

A detailed view of the instability is given in \fig{plasmoid}. 
As suggested in \cite{bulanov_78}, 
a secondary island first appears as a long-wavelength 
linear perturbation to the current sheet, 
with two $X$ points forming at the ends of the sheet [\fig{plasmoid}(b)]. 
The reconnection proceeds in a manner 
analogous to Stages I--III discussed above: as the secondary island 
grows, the two secondary $X$ points collapse, giving rise to two current 
sheets, while the island is circularized [\fig{plasmoid}(c)]. 
The primary and the secondary islands exert attracting 
forces on each other. When the secondary island is sufficiently large, 
this attraction causes it to split into two parts, which then 
coalesce with the primary island [\fig{plasmoid}(d-f)]. 
Note that the splitting of the secondary island into two is a result 
of the exact symmetry of our configuration about the $x$ axis. 
Even a slight breaking of this symmetry would cause the entire secondary 
island to move either upwards or downwards to coalesce with 
the primary \footnote{This was confirmed by L.~Chacon 
[private communication (2005)] using a grid code.}. 

We note that, given small enough $\eta$, the secondary 
current sheets should be unstable to generation of tertiary islands etc. 
Also, if the initial flux is not yet exhausted after the secondary island 
has coalesced with the primary, the primary current sheet can be 
regenerated via a second collapse (\fig{intro_figs}, Stage IVb). 
Given sufficiently large $\DD$, the cycle of current-sheet formation 
--- secondary-island generation --- coalescence may be repeated several 
times before saturation is reached. 

\begin{figure}[t]
\begin{tabular}{cc}
\psfig{file=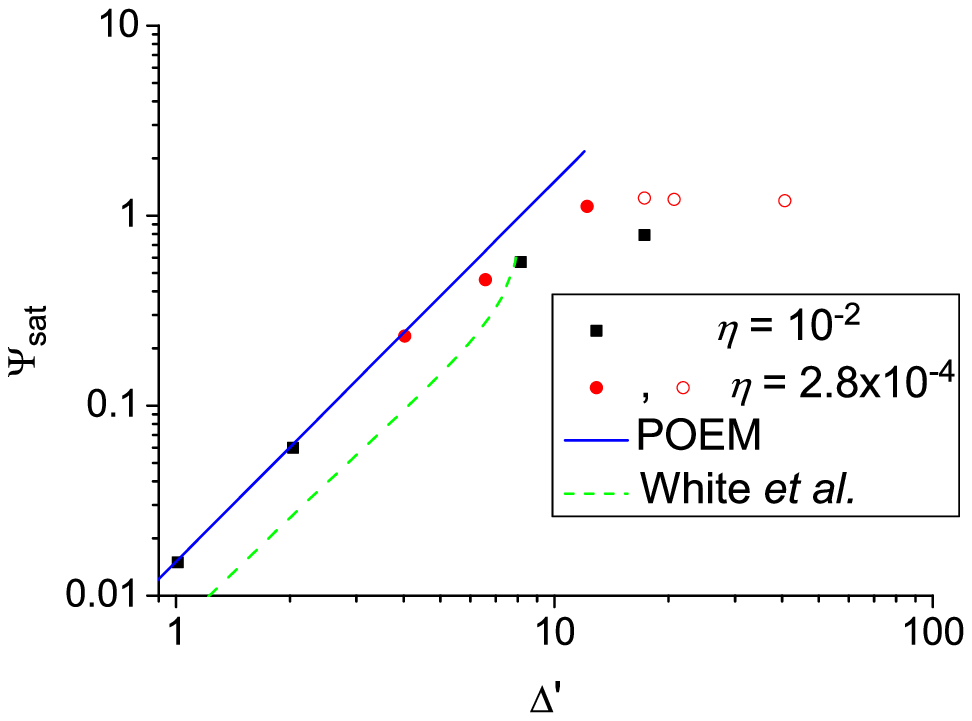,width=4.2cm} &
\psfig{file=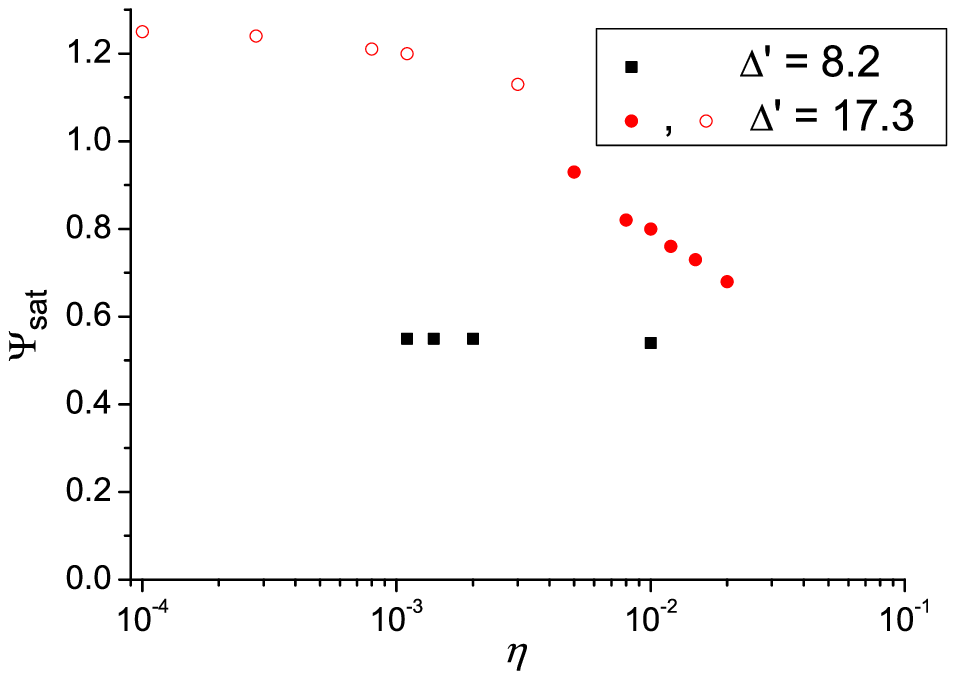,width=4.2cm}\\
(a) & (b)
\end{tabular}
\caption{\label{militello_fig} 
(a) Saturated amplitude $\psisat$ vs.~$\DD$ for different values of $\eta$. 
The theoretical curves by POEM [\eq{POEM}] and White \etal~\cite{white_77} 
are also shown. 
The island width formula~\eqref{W_vs_psi} has 
been used to convert $\Wsat$ calculated by these authors into $\psisat$. 
(b) $\psisat$ vs.~$\eta$ for $\DD=8.2,17.3$. 
In both plots, hollow points are the cases where $\Wsat$ exceeded the box size.}
\end{figure}

\paragraph{Stage V: Saturation.} 
The saturated island width in the limit of small $\DD$ 
has recently been calculated by Escande \& Ottaviani \cite{escande_04} 
and Militello \& Porcelli \cite{militello_04}, a theory henceforth referred to, 
using a liberal permutation of the first letters of the authors' surnames, 
as POEM. They found 
\be
\label{POEM}
\Wsat=2.44a^2\DD,\qquad a^2=-\psi_0''(0)/\psi_0''''(0).
\ee
For our equilibrium, $a^2=0.125$. 
\Fig{militello_fig} shows the dependence of 
the numerically obtained saturated flux on $\DD$ 
and $\eta$ compared to the quantitative predictions 
of POEM and of the earlier theory of White \etal~\cite{white_77}. 
We plot $\psisat$ instead 
of $\Wsat$ because, for the largest $\DD$ values, the island width exceeds the 
box size $L_x$ (in which case the saturation is likely to be strongly 
dependent on the equilibrium configuration). 
For $\DD\lesssim5$, there is excellent agreement 
with POEM [\eq{POEM}], but not with White \etal~\cite{white_77}. 
The occurence of the $X$-point collapse, 
i.e., whether the saturation is achieved via current-sheet 
or $X$-point reconnection, changes the saturated state: 
\fig{militello_fig} shows a jump in $\psisat$ 
at values of $\DD$ and $\eta$ for which the $X$-point collapse 
took place in Stage III. 
For sufficiently small $\eta$, the saturated amplitude does not depend on $\eta$ 
[\fig{militello_fig}(b)]. Also, $\psisat$ appears to reach a plateau 
for large $\DD$ [\fig{militello_fig}(a)], so that $\Wsat\sim$~system size. 

Note that the collapse can occur only if the saturated island width 
is larger than Waelbroek's critical value (\fig{waelb_scaling}), $\Wsat>\Wc$. 
Using \eq{Waelbroek} and \eq{POEM}, this gives $\DD\gtrsim5.2$ 
in the limit of $\eta\to0$. 

In this Letter, we have shown that, for a simple resistive model of 
the tearing mode with large $\DD$, $X$-point reconnection 
gives way to much faster current-sheet reconnection. 
Furthermore, the current sheet 
cannot exceed a certain critical aspect ratio, above which it fragments 
into secondary islands and current sheets. 
We believe the rather complex behavior we have identified 
to be a generic feature of strongly driven reconnection. 
However, a caveat is in order. While the large-$\DD$ configurations are 
often encountered in laboratory reconnection, understanding 
the physics responsible for setting up these configurations 
remains a theoretical challenge. This unknown physics, 
along with a number of kinetic effects known to be important 
in various laboratory and astrophysical contexts \cite{GEM_birn}, 
must, strictly speaking, be a part of any quantitative description of 
the tearing-mode reconnection in real plasmas. 

\begin{acknowledgments}
Discussions with J.~Drake, B.~Jemella, B.~Rogers, 
M.~Shay, and F.~Waelbroeck are gratefully acknowlegded. 
This work was supported by Funda\c{c}\~ao para a Ci\^encia 
e a Tecnologia, Portuguese Ministry for Science and Higher Education 
(N.F.L.), the UKAFF Fellowship (A.A.S.), and the 
DOE Center for Multiscale Plasma Dynamics. 
\end{acknowledgments}

\bibliography{xcollapse_PRL}

\end{document}